# Fault-Class-Matched Test Oracles for Output-Invisible Quantum Transpiler Regressions

Running title: Fault-Class-Matched Test Oracles for Quantum Transpilers

Furqan Nasir[1,2], Arif Shah[1], Iftikhar Alam[1]
[1] City University of Science and Information Technology (CUSIT), Peshawar, Pakistan
[2] National University of Computer and Emerging Sciences (FAST-NUCES), Islamabad, Pakistan.
Corresponding author: Furqan Nasir (furqannr@gmail.com)

## Abstract

Test oracles for quantum transpilers typically judge correctness by comparing compiled output against a reference: a statevector, a sampled distribution, or a unitary compared modulo global phase. A companion empirical study measures how often that choice fails. Roughly 28% of merged Qiskit transpiler bug-fixes (95% Wilson CI 19–40%) repair a fault that corrupts layout metadata, global phase, or run-to-run reproducibility while output stays correct: invisible to a black-box output-equivalence oracle by construction. This paper closes that gap with a fault-class-matched, layout-aware, width-tiered oracle family: a layout/permutation contract checker and a contract-level metamorphic relation (MR-1) for the metadata channel, a global-phase tracker for the phase channel, and a determinism runner for reproducibility. Verified from source on nine real, merged Qiskit transpiler regressions (three per channel), the output-equivalence oracle is blind throughout and the matched mechanism fires on every case. A 675-configuration sweep of the contract/metadata invariant finds no false positive. Synthetic mutant families confirm reliability at scale: 1.00 sensitivity and specificity across 36 mutants apiece for the contract/metadata and global-phase channels, and 1.00 sensitivity (95% CI 0.44–1.00) for reproducibility on the three circuits where the mutation is constructible. The contract checker costs two to six orders of magnitude less than a plain output check, the global-phase tracker is comparably cheap within its exact tier, and only the metamorphic relation carries a bounded cost. Ported natively to pytket/tket, the global-phase mechanism transfers cleanly, an identical 1.00/1.00 result with phases recovered to double-precision accuracy, evidence against a Qiskit-specific artifact.

## 1. Introduction

A quantum transpiler compiles an abstract circuit into an executable form for a specific backend, choosing a qubit layout, inserting routing swaps, and applying a sequence of optimization and rewrite passes along the way. Because a transpiler is compiler software, it can regress like any other: a bug-fix commit can repair one code path while a later change reintroduces the same class of fault, or a seemingly unrelated refactor can silently corrupt behaviour a fix depended on. Regression testing a transpiler therefore needs a test oracle, a way to decide automatically whether a given compiled circuit is correct. The near-universal choice, in both classical compiler testing and the quantum-specific literature surveyed in §3, is a black-box output-equivalence oracle: compare the transpiled circuit's execution behaviour (a statevector, a sampled measurement distribution, or a crash) against a reference, normalizing away global phase as a matter of course. A recent survey of practising quantum developers finds the same pattern outside the research literature: correctness is validated mostly by comparing against simulated output or measurement-probability distributions, and only a minority of respondents reported using any quantum-specific testing tool at all [1].

That choice has a blind spot built into its own definition rather than arising from how carefully it is applied. An output-equivalence oracle can only report a difference it can observe in a circuit's execution behaviour. A fault that corrupts something the transpiler tracks internally, such as the layout it assigns, the permutation its routing induces, the exact global phase it accumulates, or whether its own output is reproducible from one run to the next, while leaving that observable behaviour untouched, is invisible to it by construction. The companion empirical study on cross-SDK prevalence [2] measures how often this actually happens: roughly 28% of merged Qiskit transpiler bug-fixes (95% Wilson CI 19–40%) repair a fault an output-equivalence oracle cannot see, a rate that holds at 27.9% (29/104, CI 20–37%) on an extended corpus over a wider window.

Closing that gap is not a matter of running the existing oracle more carefully or on more inputs: no number of additional shots or generated programs changes an oracle's blindness to a channel its own measurement cannot represent, as this paper's own baseline measurements confirm directly (§5.4). What is needed instead is an oracle built for the specific way each fault escapes observation, and three properties follow directly from that requirement. Fault-class matching means an oracle has to target one output-invisible channel deliberately, checking the specific transpiler-internal quantity a class of fault corrupts, rather than attempting a broader, still output-based replacement for equivalence checking. Layout awareness follows from the same logic: because a transpiler's layout and routing permutation are exactly what these faults corrupt, a matched oracle has to reason about that bookkeeping directly, undoing or asserting it, rather than treat it as compiled-circuit noise to be normalized away. Width tiering, finally, means that correctness at scale cannot depend on constructing a full $2^n \times 2^n$ unitary, so a matched oracle needs an exact check for small circuits and a cheaper sampled or structural fallback beyond it, the same discipline the null output-equivalence baseline already applies to itself.

This paper builds and evaluates a family of oracles constructed on exactly those three properties. A layout/permutation contract checker and a contract-level metamorphic relation, MR-1, target the contract/metadata channel, checking invariants over the transpiler's own recorded layout and routing-permutation metadata rather than its output. A global-phase tracker targets the phase channel directly, measuring the one quantity a standard equivalence check discards by definition. A determinism runner targets the reproducibility channel, asking whether a fixed-seed, fixed-configuration transpile call returns the same result across repeated runs rather than judging a single call in isolation (§4). Each is verified from source rather than only against a synthetic proxy. §6.1 applies the full family to nine real, merged Qiskit transpiler regressions (three per channel), drawn from the same mining effort [2] reports, and finds the output-equivalence oracle blind in every case and the matched mechanism firing in every case. Synthetic mutant families built to the exact invariant each real fault violates then extend that result to scale (§5.3, §6.2). A false-positive sweep checks specificity on correct circuits (§6.3), and a cost measurement protocol prices each oracle against the same output-equivalence baseline (§5.6, §6.4). A native port of the global-phase mechanism to pytket/tket, finally, tests whether the approach travels to an independently engineered compiler (§6.5).

These pieces answer four research questions, restated here for this standalone paper. RQ1: can a matched oracle detect faults in each output-invisible channel where output equivalence is blind, on real regressions rebuilt from source? RQ2: is the family channel-specific, meaning does each oracle fire on its own channel and stay quiet on the others and on correct circuits? RQ3: what is the false-positive behaviour on correct circuits, and what does running the family cost relative to output-equivalence checking? RQ4: does the family transfer to an independently developed compiler, or is it an artifact of one SDK's own layout API? A fifth contribution falls out of the evaluation rather than being posed as a question in advance. Three data points accumulated in the course of verifying these oracles (§4.7) establish that a correctly specified detection mechanism can still miss a real fault if its trigger is not derived from the fix's own regression test, a transferable lesson about oracle engineering that extends beyond this paper's own nine faults.

The remainder of the paper is organized around these questions. §2 gives the minimal background on transpilation and the test-oracle problem this paper assumes. §3 surveys the test-oracle, differential-testing, and quantum-program-testing literature and positions this paper's contribution against it. §4 describes the oracle family and the trigger-derivation lesson. §5 sets out the evaluation design, and §6 reports results against all four research questions. §7–§10 discuss the findings, their threats to validity, their limitations, and the paper's conclusions.

## 2. Background

### 2.1 Transpilation and its metadata contracts

Qiskit [3] compiles an abstract circuit to a target backend through a staged pass manager. The transpiler assigns each logical qubit to a physical qubit through an initial layout and a final layout, both bijections from logical to physical indices, and routing inserts swaps whose net effect is a routing permutation. These are exported as metadata (TranspileLayout, final_index_layout, routing_permutation) that a downstream consumer relies on, separately from the compiled output itself.

### 2.2 The test-oracle problem for transpilation

Deciding whether a transpiled circuit is correct is an instance of the test-oracle problem [4]. When a circuit is small and unitary-pure, one can normalize the transpiled layout and compare unitaries modulo global phase. This does not scale, since an n-qubit operator has $2^n \times 2^n$ complex entries, so larger or non-unitary circuits require sampled-statevector, observable, metamorphic, or structural methods. A black-box oracle only observes the output, so a fault that corrupts internal metadata while leaving the output unitary intact can remain invisible to it by construction.

### 2.3 The three output-invisible channels (summary)

The mining study in the companion paper on cross-SDK prevalence [2] (henceforth 1A, which calls this class "equivalence-invisible" rather than this paper's "output-invisible": the same class of fault under the two papers' own terms) identifies three ways a transpiler fault can corrupt behavior while leaving the compiled output equivalent to a correct build. A contract/metadata fault corrupts the recorded layout or permutation metadata (TranspileLayout, final_index_layout, routing_permutation) without touching the executed circuit. A global-phase fault drops or double-counts a global phase that a standard output-equivalence check normalizes away by construction. A determinism fault makes a fixed-seed, fixed-configuration transpile return a different compiled representation across runs while the computation actually performed stays identical. Together these three channels account for the measured output-invisible rate reported in 1A. The three are not invisible in the same way: 1A treats contract/metadata as unconditionally invisible to any output-equivalence check, while global phase is invisible only until it is made observable (for instance under a controlled operation) and determinism only in the absence of a dedicated determinism oracle, reporting a conservative floor of 10/68 fixes (14.7%, 95% CI 8–25%) restricted to the unconditional channel alone. This paper's own global-phase tracker and determinism runner are exactly the condition-specific checks that remove that conditionality. The contract checker and MR-1 close the one channel that needed no such condition to begin with. The taxonomy and its construct-validity check are reported there and are not re-derived here.

## 3. Related Work

### 3.1 Test oracles and metamorphic testing

Judging whether a compiled circuit is correct is an instance of the test-oracle problem: exact unitary comparison does not scale past small circuits, so practice falls back on sampled, observable, metamorphic, or structural checks that trade completeness for tractability. The de-facto criterion across nearly all of this literature is a black-box output-equivalence oracle: one that looks only at the circuit's output map (a statevector, a measurement distribution, or a crash) and is blind to anything the compiler does that never surfaces there. Metamorphic testing [5] supplies one systematic way to exercise that oracle without a full specification. MorphQ [6] generates Qiskit programs and applies ten metamorphic transformations expected to preserve behaviour, then flags a fault when the transformed and original programs diverge on execution. Across more than 8,000 generated program pairs it found dozens of confirmed bugs. By the authors' own account, though, it emphasises crashes over subtler semantic miscompilation and proposes no regression-test selection or prioritisation. The same authors' more recent survey catalogues this broader quantum-software testing-and-analysis landscape in full [7]. This section synthesizes only the parts of it that bear on the output-invisible angle this paper targets. Classical software engineering has a mature, separate literature on regression test selection and prioritisation under a fixed budget (Yoo and Harman's survey [8] is the standard reference), but that literature has not yet been extended to the quantum compiler specifically. Nor does it address the question this paper is concerned with: what a selection or prioritisation scheme picks is moot if the oracle behind it cannot see the fault at all.

### 3.2 Differential and equivalence-modulo-inputs testing of compilers

A second family targets the compiler more directly by comparing its behaviour across variants, versions, or independent stacks. QDiff [9] differentially tests Qiskit, Cirq, and PyQuil by running semantically equivalent circuit variants across each and comparing the resulting output distributions statistically (Kolmogorov–Smirnov tests, cross-entropy) on both simulators and IBM hardware. Separating genuine software faults from hardware noise and sampling variation is the technique's own acknowledged difficulty. QEMI [10] adapts equivalence-modulo-inputs to the quantum setting by generating variants that remove quantum-control-flow dead code, then flags a bug when one variant crashes while the other does not, or their output distributions differ significantly. Its authors describe it as the closest of the family to regression testing proper, since variants can be reused across compiler versions, though still without a selection or prioritisation mechanism. QuteFuzz [11] fuzzes quantum compilers with structurally generated circuits (including control flow and nested subcircuits) run through multiple stacks (pytket, Qiskit, Cirq) to surface crashes, cross-stack inconsistencies, and silent miscompilations. Most recently, QSPE [12] enumerates skeletal quantum programs up to α-equivalence and validates them by statevector comparison rather than measurement sampling, reporting 708 miscompilations across quantum libraries, 81 of them acknowledged by the Qiskit team. That is a concrete demonstration that differential testing at this scale still finds real compiler faults. Equivalence-checking techniques that compare circuits directly before and after compilation (Burgholzer and Wille [13]) sit alongside this family as a more targeted, non-generative alternative. Every technique in this group, whatever its generation strategy, adjudicates a fault the same way: by comparing some projection of program output (a statevector, a sampled distribution, a crash) across two runs. None inspects the compiler's internal metadata or layout bookkeeping directly.

### 3.3 Quantum program testing and verification

A third body of work tests or verifies quantum programs and compiler passes by other means. QuCheck [14] brings property-based testing to Qiskit, letting developers state properties and generators the way Hypothesis or QuickCheck would for classical code. Correctness is still adjudicated by executing the generated programs and checking the stated property against the observed output. Mutation testing [15] supplies a complementary angle by asking not whether a fault exists but whether an existing test suite would catch one: Muskit [16] applies add/remove/replace-gate operators across nineteen supported gate types (57 operators in total) to generate mutants, and reports a mutation score computed under either a Wrong-Output Oracle or an Output-Probability Oracle, both output-based measures. QMutPy [17] combines twenty classical MutPy operators with five quantum-specific ones (gate replacement, deletion, insertion; measurement insertion, deletion). On 24 real Qiskit programs, it found that existing test suites killed only 325 of 696 generated mutants (46.7%). That is evidence that even where quantum programs have tests, those tests miss many introducible faults. Which oracle would close that gap is a question QMutPy leaves open. LintQ [18] takes a different, non-execution-based approach entirely: ten static analyses built on quantum-specific program abstractions flag patterns such as operating on a corrupted quantum state, a redundant measurement, or an incorrect sub-circuit composition, reaching 91% precision on 7,568 real Qiskit programs. Because it never runs the compiled circuit, LintQ is not subject to the output-equivalence oracle's blind spot at all. By the same token, though, it verifies source-level program patterns, not that a specific compiler transformation preserved a specific circuit's semantics, so it does not answer the question this paper asks. Formal per-pass verification (Giallar [19] and CertiQ [20]) is the one family that inspects the transformation itself rather than only its output, at the cost of a substantial proof burden and partial pass coverage (44 of 56, and 26 of 30, Qiskit passes respectively). §5.4 revisits both quantitatively as the closest existing alternative to this paper's own family. Bugs4Q [21], finally, is a benchmark rather than a technique (42 real, manually validated Qiskit bugs collected from GitHub, Stack Overflow, and Stack Exchange), useful as a reusable regression corpus but silent on what kind of oracle should be run against it.

### 3.4 Positioning

Read together, these three literatures share a structural property this paper turns into its central argument. Whether a technique differentially tests, metamorphically transforms, fuzzes, mutates, or property-checks a quantum program, its verdict is computed from some projection of program output (a statevector, a sampled measurement distribution, a crash, or, in mutation testing, whether an output-based test suite's assertions still pass). LintQ [18] is the sole exception, and only because it inspects source code and never executes anything. It therefore cannot verify that a specific compiler pass preserved a specific circuit's semantics, which is the question a regression check on the transpiler has to answer. Formal per-pass verification (Giallar [19], CertiQ [20]) is the only family that inspects the transformation itself, but neither is a lightweight runtime check a CI pipeline could run per fix, and both cover a minority of passes by their own account (§5.4). No technique in this survey is constructed to detect a fault that is provably invisible to output comparison, because, as §2.3 argues and §6 confirms empirically, that is a real and populated category, not a corner case. Metamorphic testing comes closest in spirit: MorphQ's [6] metamorphic relations state what should stay invariant under a transformation of the program, checked by re-executing it. This paper's contract-level metamorphic relation (MR-1, §4.3) applies the same idea one level down, to what should stay invariant in the compiler's own metadata (a qubit permutation, a routing contract) under a transformation of the circuit, checked without executing anything at all. That shift, from a metamorphic relation over program behaviour to one over compiler bookkeeping, together with a global-phase tracker and a plain contract checker targeting the other two output-invisible channels (§2.3), is what this paper's family contributes that the surveyed literature does not. The result is a set of oracles deliberately built to see past the output-equivalence oracle's own observation boundary, rather than to exercise that boundary more thoroughly.

## 4. The Oracle Family

### 4.1 Design principles

Fault-class matching: each oracle targets exactly one output-invisible channel rather than attempting a general-purpose replacement for output equivalence. Layout awareness: every oracle reasons about the transpiler's layout/permutation contract rather than the raw output alone. Width tiering: an exact operator only for small circuits ($n \leq 12$ qubits), sampled layout-normalized statevectors for 13–22 qubits, and a structural fallback beyond. No oracle ever materializes a full $2^n \times 2^n$ unitary at scale. One exception is stated here rather than left for the cost results to reveal: MR-1 (§4.3) has no sampled-tier fallback of its own and is restricted to its own $n \leq 12$ precondition, because the permutation gate its check appends has no small local factor to exploit past that width. §6.4 reports what that restriction costs in practice.

### 4.2 Contract checker (contract/metadata channel)

Asserts the transpiler's layout and permutation contract: the initial and final index layouts must be valid permutations, and they must compose consistently with the routing permutation (final = routing ∘ initial).

### 4.3 Contract-level metamorphic relation MR-1 (contract/metadata channel)

Following the metamorphic-testing paradigm [5], MR-1 (permutation consistency) requires that appending the recorded routing_permutation to the routed output recover the input unitary. A violation signals a contract fault the output oracle cannot see, since the layout-applied output can remain correct even when the recorded permutation is wrong.

### 4.4 Global-phase tracker (global-phase channel)

The exact complement of the output-equivalence oracle: where the latter compares modulo global phase, this oracle measures the phase itself, so a dropped or double-counted global phase is detected end to end. Global phase merits a dedicated oracle even though a standalone circuit measured in the computational basis is insensitive to it, because the phase becomes physically observable the moment the circuit is used as a controlled subroutine or composed interferometrically.

### 4.5 Isolated-pass exception differential (auxiliary mechanism)

The isolated-pass exception differential targets the same contract/metadata channel from a different angle, developed for #14603 specifically. In the full pipeline, a downstream pass recomputes the permutation metadata that the buggy ElidePermutations corrupts, so both the output oracle and a full-pipeline property comparison stay blind on both the fixed and buggy build. The fault is invisible to any check run after the rest of the pass manager has had a chance to paper over it. The mechanism instead runs ElidePermutations alone, outside the rest of the pass manager, against a trigger built from the fault's own regression test, once against a from-source build of the fixed revision and once against the buggy revision. It then compares two things: the pass's own output circuit and the property_set entry it records (virtual_permutation_layout). A divergence in either is the signal. In the #14603 case, the buggy pass instead raises an exception where the fixed pass completes cleanly, an asymmetric failure that the downstream pass's silent recomputation would otherwise erase before the full-pipeline output oracle ever ran.

### 4.6 Determinism runner (determinism channel)

The determinism runner targets fixed-seed reproducibility rather than a single transpile call. A minimal reproduction first surfaced the channel independently of any historical fix: with a fixed seed_transpiler and an unchanged circuit, coupling map, and basis, repeated transpile() calls on Qiskit 2.4.2 and 2.5.0 return different final_index_layout metadata across otherwise identical calls. This non-determinism is present even single-threaded, traced to the time-budgeted VF2 layout search, while the compiled circuit itself stays functionally correct. The runner generalizes this into a source-verification protocol for #14730 and #16237. A trigger circuit taken from the fix's own regression test is transpiled with a fixed seed_transpiler across several PYTHONHASHSEED values and several repeats per value, subprocess-isolated, since PYTHONHASHSEED is fixed for the lifetime of one interpreter and cannot be varied mid-process. Each run is then fingerprinted twice: a RAW fingerprint (the exact compiled instruction sequence and recorded layout metadata) and a FUNCTIONAL fingerprint (the sorted statevector probability vector, invariant to qubit permutation). A fixed build with raw_distinct = 1 alongside a parent with raw_distinct > 1, while func_distinct stays 1 on both, is the signature this study treats as source-evidenced, output-invisible determinism, the same computation compiled differently across runs. #14730 shows exactly this pattern (fix raw_distinct = 1, parent raw_distinct = 6 across 5 PYTHONHASHSEED values × 10 runs). #16237 shows the same pattern one layer down, in ConsolidateBlocks' choice of equivalent basis-gate name (rzz/cz on the fix at every run; ryy/rzz or ecr/cx/cz on the parent depending on hash seed, both consolidations semantically equivalent).

### 4.7 Trigger derivation: the detection mechanism and the trigger are a matched pair

A trigger that looks plausible from a PR's title or description is not enough to make a detection mechanism fire: the mechanism and the trigger that exercises it are a matched pair, and this study accumulated three data points establishing that the hard way. The first came from #14939, a further contract/metadata fix beyond H1 and H2: probing the contract checker with a generic full-pipeline trigger produced no visible property divergence on either build, establishing only that the generic check is blind and leaving a targeted isolated-pass confirmation outstanding. The second came from an initial attempt at #15024 (register preservation in TranspileLayout). The trigger compared the per-virtual-bit register membership of two plain QuantumRegisters between a fixed and a buggy build, and came back negative on both output and mechanism: not because the fault was absent, but because the compared field is derived from the circuit's own qubits and is therefore identical on both builds by construction, regardless of what the layout itself records. Reading the fix's own regression test (test_layout_registers_preserved) showed what the trigger needed to be: a circuit built from an AncillaRegister together with a QuantumRegister, transpiled with an explicit initial_layout, comparing initial_layout.get_registers() rather than per-qubit membership, swept across optimization levels 0–3. Rebuilt this way, the same trigger went 4/4 positive: the buggy build drops the register set entirely while the fixed build preserves it, at every optimization level, while the output-equivalence oracle stays blind at every level too. The mechanism had been correct throughout. Only the trigger was wrong. Three data points now support the same conclusion (#14939, #15024 v1, #15024 v2): a detection mechanism's specification is necessary but not sufficient, and a targeted trigger has to be derived from the fix's own regression test rather than invented from the PR's title or description. The #15024 v2 result is also reported in its own right, as a source-evidenced fault in Table 4 (§6.1). The same discipline succeeded twice more under the same standard: #16215 (Commuting2qGateRouter) and #15040 (DAG edge-order) were each confirmed with a trigger taken directly from its own fix commit's regression test rather than a generic circuit family, the eighth and ninth source-evidenced faults added to Table 4 in that order.

## 5. Evaluation Design

This section sets out the evaluation design that follows from that gap: an event model and cohort design (§5.1), fault and mutant construction (§5.2-§5.3), comparator baselines (§5.4), validity gates and claim scope (§5.5), and a cost measurement protocol (§5.6). Figure 1 gives the pipeline these subsections instantiate, from a static manifest and instrumented transpilation through change-event construction, width-tiered oracle application, and metric computation, all resting on a shared reproducibility substrate.

### 5.1 From-source event model and cohorts

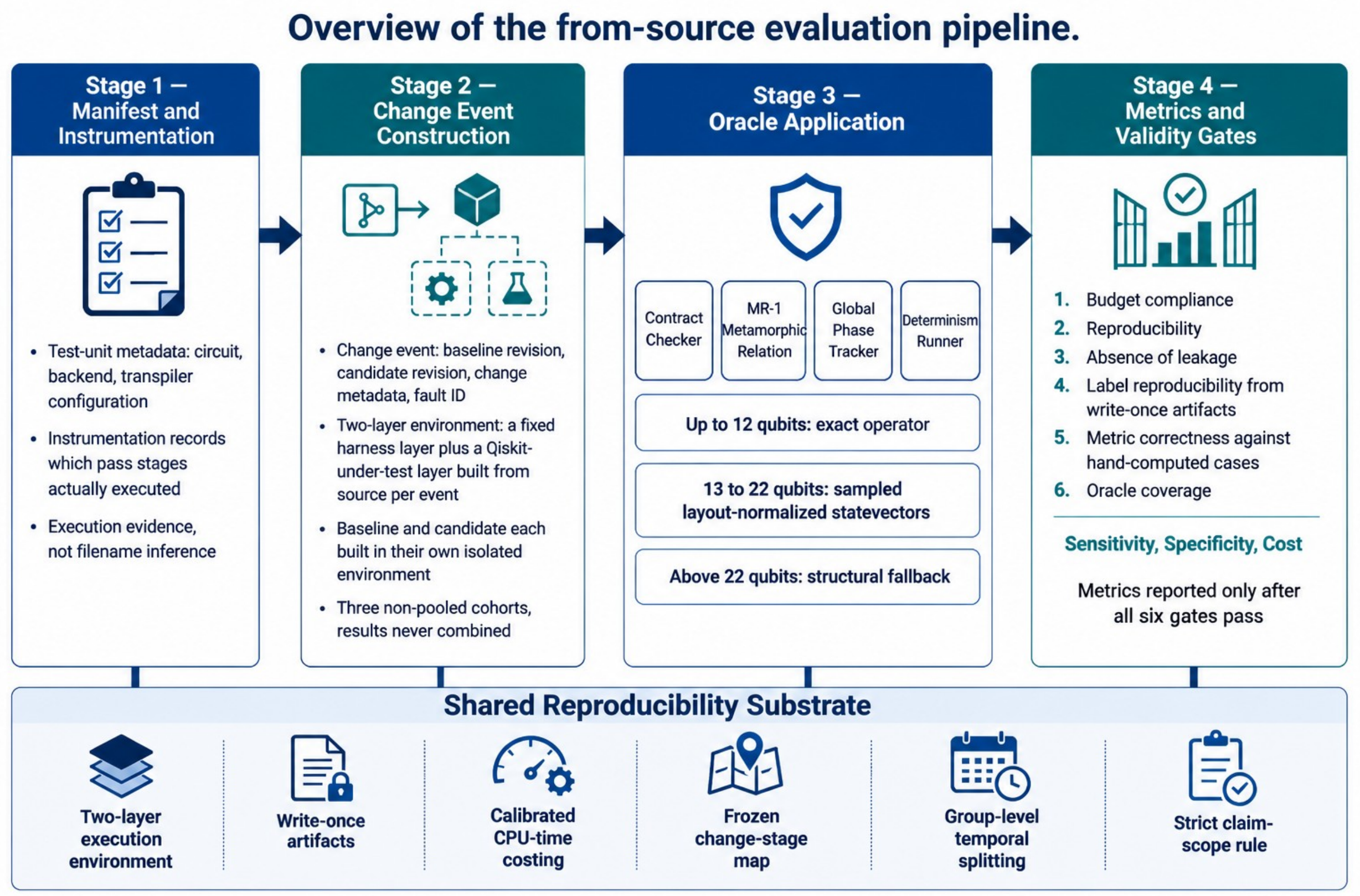

**Figure 1. Overview of the from-source evaluation pipeline: a static manifest and instrumented transpilation, change-event construction under non-pooled cohorts, width-tiered oracle application, and metric computation under six validity gates, resting throughout on a shared reproducibility substrate (a two-layer execution environment, write-once artifacts, calibrated CPU-time costing, a frozen change-stage map, group-level temporal splitting, and a strict claim-scope rule).**

The unit of the dataset is a change event: a tuple (baseline_revision, candidate_revision, change_metadata, fault_id). Every event declares one of three cohorts (Figure 2), reported separately and never pooled. A forward_regression event pairs the last known-good parent with the regression-inducing commit, and is the only cohort that models the real CI question. A fix_boundary_differential event is used in reverse orientation when the introducing commit has not been traced: it pairs a fix commit (baseline) with its parent (candidate). A mutation event injects a single controlled fault on a base revision.

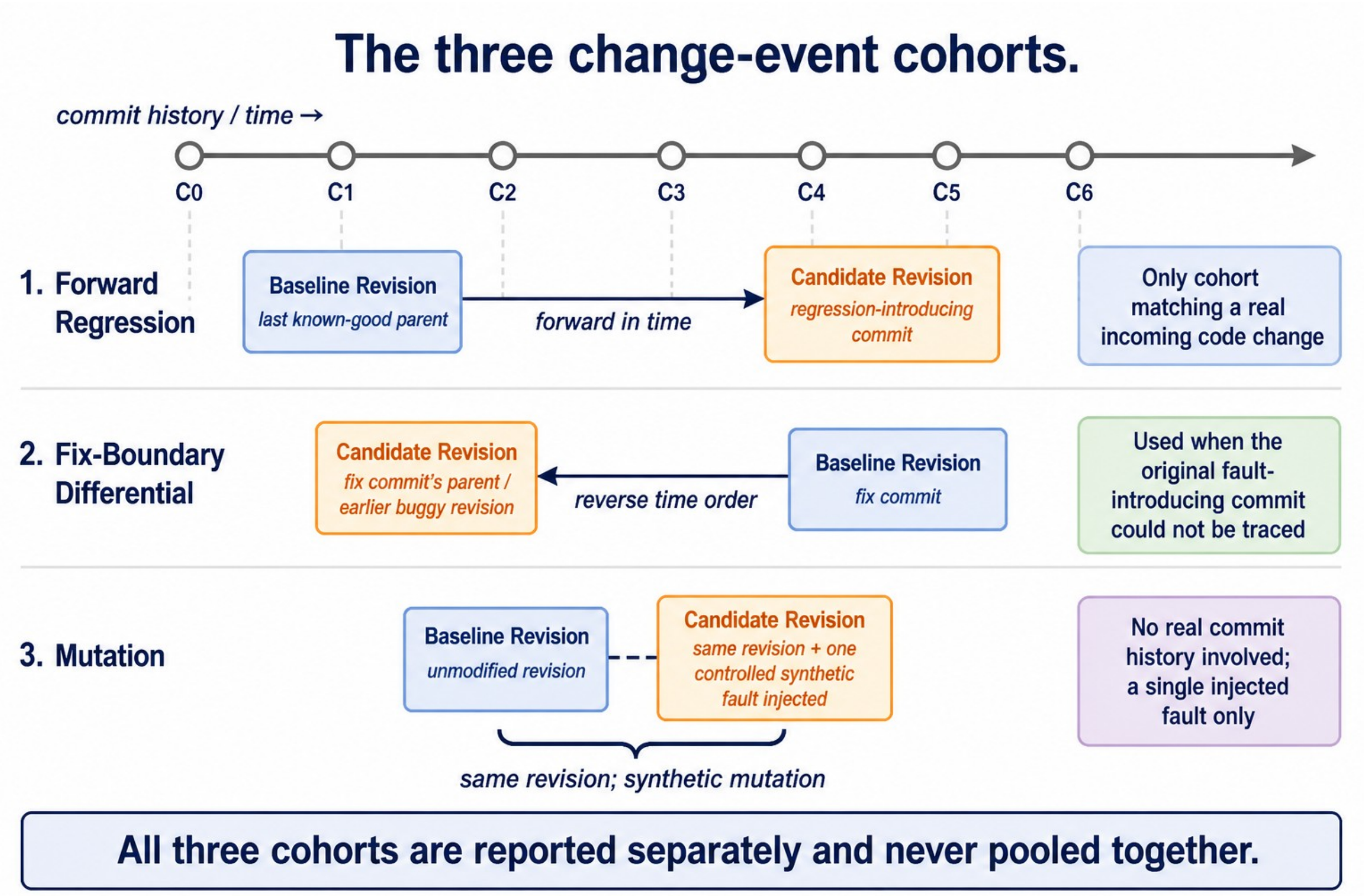

**Figure 2. The three change-event cohorts (forward regression, fix-boundary differential, mutation) and how each pairs a baseline revision with a candidate.**

The five historical events (Table 1) anchor the from-source reproduction. The definitive provenance ledger (Table 2) records exact build SHAs and how each introducing commit was established.

**Table 1. Summary of the five historical events H1–H5, ordered H1–H5 (full provenance in Table 2, cohorts never pooled).**

| Event(s) | Cohort | Fault type | Status |
|---|---|---|---|
| H1 ElidePermutations (#14603) | fix_boundary | contract/metadata | source-evidenced, output-invisible |
| H2 final_layout composition (#14919) | fix_boundary | contract/metadata | source-evidenced, output-invisible |
| H3 VF2Layout determinism (#14730) | fix_boundary | determinism | source-evidenced, parent non-deterministic |
| H4 VF2PostLayout no-op (#14120) | forward regression | performance | confirmed (Stage-2 multi-run, up to ~316x at 27q) |
| H5 VF2Layout panic (#16285) | fix_boundary | functional | blocked (not run) |

**Table 2. Definitive event-provenance ledger. Good is the fixed revision for fix_boundary events and the last-good parent for the forward_regression event (H4). Buggy is the parent for fix_boundary events and the introducing candidate for H4. Introducing-commit provenance is PR-reference reading rather than automated bisection: no bisection run for this study ever converged on a failing probe, so none of these commits should be described as bisection-traced.**

| Event (PR) | Channel, cohort, detection | good ↔ buggy (SHA) | Introducing commit | Trigger provenance | Status |
|---|---|---|---|---|---|
| H1 (#14603) | contract/metadata, fix_boundary, retrospective | 96fda188 ↔ 7c3890da | #13094 (cbb4d5d5), PR-referenced | extracted from fix test | source-evidenced, output-invisible |
| H2 (#14919) | contract/metadata, fix_boundary, retrospective | dfcc5c6c ↔ 14df5941 | #11399 (df59ab0c), PR-referenced | synthesized (identity) | source-evidenced, output-invisible |
| H3 (#14730) | determinism, fix_boundary, retrospective | d33ef533 ↔ 056c6413 | not traced | extracted from fix test | source-evidenced, parent non-deterministic |
| H4 (#14120) | performance, forward_regression, prospective | a18e1516 ↔ a8d23667 | a8d23667 (candidate) | synthesized (GHZ) | confirmed Stage-2, up to ~316x at 27q |
| H5 (#16285) | compilation_failure, fix_boundary, retrospective | 0b2cdf2b ↔ f0fae6f3 | not traced | extracted from fix test | not run |

## 5.2 Evidence grading (Level 1-3)

Level 1 (built and executed): both revisions are built from source and the fault is triggered by the specified detection mechanism while the output oracle stays blind. Level 2 (fix-code and regression-test inspection): the fix commit and its own regression test are read to confirm the coded channel. Level 3 (dedicated-runner reproduction): a released runner re-executes the fault from source across the varying condition. Levels 1 and 3 are the strongest, executable evidence. Level 2 is corroborating.

## 5.3 Mutant populations per channel

Two mutant families were built to extend the global-phase family's sensitivity/specificity evidence (§6.2) to the remaining two channels, on the same six trigger circuits (GHZ and QFT, 4–6 qubits). The contract/metadata family applies one of six permutation-valid corruption modes (pairwise swaps, cyclic rotations, a full reversal) to the transpiled circuit's recorded routing_permutation metadata, leaving every compiled instruction and the initial/final index layouts untouched. Because each mode is itself a bijection of range(n), the plain permutation-validity check never fires, isolating detection to the same layout-composition inconsistency #14603 and #14919 exhibit. The determinism family reproduces the mechanism behind #14730 and #16237 directly rather than approximating it: after an otherwise fully seeded transpile, one adjacent pair of instructions acting on disjoint qubits (which therefore commute trivially, so reordering them cannot change the circuit's unitary) is kept or swapped according to next(iter({"keep", "swap"})), a two-element string set whose iteration order is exactly the PYTHONHASHSEED-dependent tie-break believed to drive #16237's basis-gate selection. Neither family reuses the mutation engine built for the companion cost-aware test-selection study. Both are purpose-built for their channel. Because each family's mutation operator is defined as exactly the invariant its own channel's oracle checks, a positive result on these mutants shows that the mechanism catches a violation of that invariant reliably at scale. It does not by itself establish that the invariant is the right one to check in the first place. That construct-validity question is answered independently by the nine real, source-verified faults in §6.1, not by the mutants built around them.

## 5.4 Baselines

Comparators are identified from the related-work survey in §3, built by following citations and venues across the test-oracle, differential-testing, and quantum-program-testing literature. That survey identifies six testing techniques and one real-bug benchmark as the field's established alternatives to a matched-oracle approach, listed in Table 3 together with the oracle mechanism each declares and whether that mechanism can in principle observe the contract/metadata, global-phase, or determinism channel this study targets. A recent methodological review of 59 empirical quantum-software-testing studies examines baseline-comparison practice as one of its central dimensions and reports it applied inconsistently across the sampled literature [22]. The explicit, mechanism-by-mechanism comparator analysis in Table 3 is this paper's own answer to that gap.

**Table 3. Testing techniques identified in §3's related-work survey as comparators, their declared correctness oracle, and whether that oracle can in principle observe the contract/metadata, global-phase, or determinism channel this study targets. The "No" verdicts for QDiff, MorphQ, QuteFuzz, and QEMI are analytically derived from each tool's published oracle design; none of the four was re-executed against this study's faults. check_semantic's own blindness on the same nine faults and every mutant is measured directly (Tables 4-5), not inferred.**

| Study | Oracle mechanism | Channel access | Basis for verdict |
|---|---|---|---|
| QDiff (Wang et al., 2021) | Statistical comparison of output distributions across quantum-software stacks | No, on all three channels | Output-based; provably invariant on this study's contract, phase, and determinism mutants (§6.2) |
| MorphQ (Paltenghi & Pradel, 2023) | Metamorphic relations between original and transformed program outputs | No, on all three channels | Same output-based invariant |
| QuteFuzz (Iwumbwe et al., 2025) | Crash and cross-stack compiler/simulator inconsistency detection | No, on the silent, non-crashing faults this study targets | All nine real faults and every synthetic mutant compile and execute cleanly; QuteFuzz's own reported weakness is silent-miscompilation detection |
| QEMI (Luo et al., 2026) | Equivalence modulo inputs: crash or output-distribution difference between a program and its variant | No, on all three channels | Same output-based invariant |
| Giallar (Tao et al., 2022) | Formal per-pass semantic-preservation proof (symbolic execution, SMT, rewrite rules) | Potentially, for a covered pass and property | Could in principle catch a contract-level fault if that pass is proved; 44/56 Qiskit passes covered, no runtime check on an unproven build |
| CertiQ (Shi et al., 2020) | Formal per-pass verification via contracts and SMT | Potentially, same caveat as Giallar | 26/30 passes reimplemented and verified; proof/specification burden, limited public reproducibility |

The first four techniques share one property regardless of how each generates or transforms programs: their oracle is a function of execution output alone, whether a statistical comparison of measurement-outcome distributions (QDiff), a metamorphic relation between output behaviours (MorphQ), a cross-stack execution inconsistency (QuteFuzz), or an equivalence-modulo-inputs output/crash comparison (QEMI). That is exactly the oracle class this study's own null baseline instantiates: check_semantic, the output-equivalence oracle implementation used as this paper's baseline throughout. Tables 4 and 5 already measure its sensitivity on every fault and mutant in the study: blind on 9/9 real faults, and blind on every contract mutant, every phase mutant, and every constructible determinism circuit. The zero generalises to all four techniques by more than analogy. A contract-channel mutant corrupts only recorded layout metadata and never touches an executed instruction, so the circuit any of these tools would actually run is bit-identical whether or not the metadata is corrupted. A global-phase offset is unobservable under any computational-basis measurement, sampled or exact, because measurement probabilities depend on amplitude magnitude and cancel the phase exactly. The determinism mutant, finally, is constructed to leave the output probability distribution unchanged, confirmed directly by func_distinct = 1 on every trial (§6.2). No number of additional shots, transformations, or generated programs changes an oracle's blindness to a channel its own measurement cannot represent.

Giallar and CertiQ sit in a different part of the design space. Both verify semantic preservation of an individual compiler pass directly, by symbolic execution and SMT rather than by comparing execution outputs, so a contract-level fault could in principle be caught if the specific pass and property happen to be covered by an existing proof. In practice that coverage is partial (44 of 56 passes for Giallar, 26 of 30 for CertiQ), and checking Giallar's own enumerated pass list against this study's faults is informative: ElidePermutations, the pass behind #14603, does not appear among its 44 verified passes, nor anywhere in the 56-pass universe Giallar reports, so this specific fault sits outside what that proof could have caught irrespective of the coverage percentage. The guarantee is also per-pass and static: it says nothing about a build whose passes have not been (re)proved, and it cannot run at CI time against an arbitrary compiled circuit the way a lightweight runtime oracle can. This is the gap the contract checker, MR-1, and the global-phase tracker fill: no proof obligation per pass, applicable to the full pass manager rather than a verified subset, and cheap enough to run on every build (§5.6).

Bugs4Q [21] is a reusable corpus of 42 real, manually validated Qiskit bugs rather than a testing technique, so it is not a comparator here; it is a candidate source for future retrospective breadth studies (§6.6). QuCheck [14], the property-based testing framework for Qiskit discussed in §3.3, is likewise not included in Table 3's comparison: its correctness check is still adjudicated by executing the generated program and inspecting the observed output, the same output-based class already covered by the four comparators Table 3 does include.

### 5.5 Validity gates and claim scope

Six gates check implementation correctness before any result is reported: budget compliance, reproducibility, absence of leakage, label reproducibility from write-once artifacts, metric correctness against hand-computed cases, and oracle coverage. A claim-scope rule forbids any headline temporal-generalization claim while fewer than three forward_regression events are verified in the held-out cohort. Results are then reported per event. The verified forward-regression cohort is 1 (H4 #14120, confirmed). #15024 is fix_boundary_differential evidence (Table 4), not a forward_regression event, and is not counted in this cohort. See data/events/PROVENANCE_BACKLOG.md.

### 5.6 Cost measurement protocol

Each oracle's per-call CPU cost is measured against check_semantic, the plain output-equivalence check this study otherwise treats as the null baseline, on GHZ circuits swept across n = 4, 8, 12, 16, 20 qubits. The sweep spans the family's own exact tier ($n \leq 12$, a full 2^n × 2^n operator is permitted) and its sampled tier ($12 < n \leq 22$, statevector probes only, no full operator). Timing follows a fixed calibration discipline used throughout this project: CPU process_time, thread-pinned (RAYON_NUM_THREADS=1, OMP_NUM_THREADS=1, set before importing Qiskit), one discarded warmup call, and the median of five repeats. time.process_time() on the measurement machine (Windows) is quantized to the OS scheduler's ≈15.6 ms CPU-accounting tick, so each repeat times a calibrated batch of back-to-back calls sized to clear 500 ms in total and divides to recover the per-call cost, the same averaging technique timeit uses. A call already expensive enough to clear that target alone (every n = 12 exact-tier call) is timed unbatched. Hardware: Windows 10 (build 26200), Intel Family 6 Model 141 Stepping 1 (AMD64, 16 logical cores), Python 3.11.9, RAYON_NUM_THREADS = OMP_NUM_THREADS = 1.

## 6. Results

### 6.1 RQ1: nine source-evidenced faults across three channels

Channel labels and their construct-validity check are reported in the companion prevalence study (1A), whose own source validation independently confirms the coded channel for all nine faults examined below (1A Table 2: 16 of 68 mined fixes checked directly against their own source diff and regression test, with no label overturned). The channel a fault belongs to is therefore established by 1A's independent mining and adjudication protocol, not assigned by the detection mechanisms built here. Here we treat the nine as given and evaluate detection. We verify nine of the mined output-invisible fixes directly from source, three per channel, applying the detection mechanism specified for each.

**Table 4. The nine source-evidenced faults as detection targets: output-equivalence-oracle verdict vs. matched-mechanism verdict (three per channel). The output-equivalence oracle is blind in every case. The channel-matched mechanism fires in every case.**

| Fault (PR) | Channel | Evidence level | Output-equivalence oracle | Matched mechanism |
|---|---|---|---|---|
| #14603 | contract / metadata | Level 1: built and executed | Blind: compiled output equivalence unaffected | Layout/permutation contract checker fires (isolated-pass exception differential) |
| #14919 | contract / metadata | Level 1: built and executed | Blind | Contract-level metamorphic relation MR-1 fires (permutation-consistency violation) |
| #15024 | contract / metadata | Level 3: dedicated runner (4 optimization levels: 0–3) | Blind: compiled output equivalent at every optimization level | Layout/permutation contract checker fires: initial_layout.get_registers() empty on the buggy build vs. the full register set on the fix, at every level |
| #14956 | global phase | Level 1: built and executed | Blind | Global-phase tracker fires |
| #16201 | global phase | Level 3: dedicated runner | Blind: the two unitary matrices are identical modulo phase (u_re/u_im blocks match) | Global-phase tracker fires: fix global_phase = 1.3464 vs. parent global_phase = 1.7952 |
| #16215 | global phase | Level 1: built and executed (isolated-pass differential) | Blind: routed circuit equivalent to the original modulo phase | Global-phase tracker fires: fix global_phase = 0.3 (preserved) vs. parent global_phase = 0.6 (corrupted) on a circuit with no commuting blocks |
| #14730 | determinism | Level 3: dedicated runner (50 runs: 5 PYTHONHASH SEED values × 10 repeats) | Blind: func_distinct = 1 on both fix and parent (functionally identical every run) | Determinism runner fires: raw_distinct = 1 (fix) vs. raw_distinct = 6 (parent, across seeds) |
| #16237 | determinism | Level 3: dedicated runner (6 repeated runs) | Blind: compiled circuit functionally correct on both builds | Basis-gate selection inconsistent on the parent (ryy/rzz, ecr/cx/cz vary run to run) vs. consistently rzz/cz on the fix |
| #15040 | determinism | Level 3: dedicated runner (10 independent process restarts) | Blind by construction: the trigger is a Barrier, which alters no circuit semantics on either build | DAG edge order diverges: 0/10 structurally-inconsistent restarts on the fix vs. 10/10 on the parent (DAGCircuit.structurally_equal on independently rebuilt DAGs) |

On nine real, merged Qiskit transpiler regressions (three per channel, rebuilt or re-executed from source), the black-box output-equivalence oracle is blind in every case, while the channel-matched detection mechanism fires in every case. Four (#14603, #14919, #14956, #16215) are verified by building both revisions from source and running the full detection pipeline (Level 1). The remaining five (#14730, #16237, #16201, #15024, #15040) are verified by a dedicated runner re-executing the fault from source across a varying condition (Level 3). #14730's runner sweeps 50 runs (5 PYTHONHASHSEED values × 10 repeats), #16237's sweeps 6 repeated basis-selection runs, #16201 is confirmed by direct comparison of the two builds' compiled global phase, #15024 by comparing initial_layout.get_registers() between the two builds across optimization levels 0–3, and #15040 by comparing DAGCircuit.structurally_equal() across 10 independent process restarts per build. #16215 and #15040 were confirmed the same way #15024 was, by rebuilding the trigger from the fix's own regression test (§4.7) rather than the generic circuit family the project's own preliminary #16215 script had used and which never reached the buggy code path. #15040 is reported with one caveat the others do not need: its observable is the compiled DAG's internal edge order, not the compiled circuit's execution behaviour, and its trigger (a Barrier) is semantically a no-op on either build, so "output-invisible" here is definitional rather than an emergent property of a specific compiler transformation. It is included as a real, source-verified instance of the same failure mode this study's other determinism faults exhibit, not silently equated with #14730 and #16237's transpile-level evidence.

One clarification against the underlying event ledger: H1 (#14603) is recorded there with a cohort-placement status of "rejected" from the primary, production-pipeline cohort. This is not a detection-outcome status: H1 was moved to the secondary retrospective evaluation specifically because confirming it requires the property-level, isolated-pass harness reported above, which departs from the standard transpile-and-oracle pipeline. The Level-1 built-and-executed detection result in Table 4 holds regardless. A reader consulting the raw ledger should not read that cohort-placement label as evidence the mechanism failed to fire.

A tenth candidate is worth naming rather than passing over silently. 1A's own source validation independently confirms #16402 as a genuine global-phase fault. A trigger built from its fix commit's own regression test, a CommutativeCancellation pass accumulating a PhaseGate(π/4) against fifteen T gates, was run against both revisions and did not reproduce it: the tracked global phase came back 0.0 on the fixed build and the buggy parent alike. Per §4.7's own discipline, a negative result from one trigger shape is inconclusive, not disconfirming, since the fault may simply not be reachable through this specific accumulated-rotation count. #16402 is accordingly left out of Table 4 rather than force-fit into it, pending a further attempt with a different gate or repeat count drawn from the same regression test.

## 6.2 RQ2: channel specificity

A family of 36 synthetic global-phase faults, six phase offsets across six circuits, is detected with sensitivity 1.00 (95% Wilson [23] CI 0.90–1.00) while the output oracle stays blind, at specificity 1.00 (95% CI 0.80–1.00) on the clean baselines. A matched-oracle call runs in a median of 17.6 ms with 383 KB of peak traced memory.

The contract/metadata family shows the same shape: 36 permutation-valid routing_permutation corruptions (six corruption modes across the same six circuits) are detected with sensitivity 1.00 (95% Wilson CI 0.90–1.00), and the output-equivalence oracle stays blind on every one. The six clean baselines show specificity 1.00 (95% CI 0.61–1.00), with zero false positives and zero spurious global-phase firings, confirming the mutation does not leak into a different channel.

The determinism family reorders one adjacent, qubit-disjoint (hence trivially commuting) instruction pair according to a PYTHONHASHSEED-dependent tie-break, under the same protocol used to source-verify #14730 and #16237 (§6.1): six hash seeds × five repeats per circuit. Across the three QFT circuits, every run shows exactly two distinct raw fingerprints against one functional fingerprint (raw_distinct = 2, func_distinct = 1), the same output-invisible non-determinism signature as the two real faults, giving sensitivity 1.00 (95% CI 0.44–1.00; n = 3). The three GHZ circuits offer no eligible instruction pair under a line coupling map (their compiled output never places two disjoint-qubit instructions adjacently), so the mutation is a structural no-op there rather than a missed detection. Sensitivity is reported over the three circuits where the fault is constructible, mirroring MR-1's own GHZ-only precondition earlier in this table. At n = 3 the 95% Wilson interval is necessarily wide (0.44–1.00), a direct consequence of the small circuit count rather than of any observed failure: all three constructible circuits registered a positive detection, and the interval should be read as consistent with high sensitivity rather than as a tight estimate of it. All six circuits show the unmutated control as raw-stable across every hash seed (specificity 1.00, 95% CI 0.61–1.00, 6/6).

**Table 5. Held-out channel-matched mutant-family metrics. Sensitivity is the fraction of mutants each channel's oracle(s) detect while the output oracle stays blind (for determinism, the denominator is the three circuits where the fault is constructible, see text). Specificity is the fraction of clean baselines on which the oracle does not fire. Both with 95% Wilson intervals.**

| Channel | Mutation family | Sensitivity (95% CI) | Specificity (95% CI) |
|---|---|---|---|
| global phase | 36 synthetic phase mutants (6 offsets × 6 circuits) | 1.00 (0.90–1.00) | 1.00 (0.80–1.00) |
| contract / metadata | 36 routing-permutation corruptions (6 modes × 6 circuits) | 1.00 (0.90–1.00) | 1.00 (0.61–1.00) |
| determinism | PYTHONHASHSEED reorder, 3/3 constructible circuits (QFT4-6; n/a on GHZ) | 1.00 (0.44–1.00, n=3) | 1.00 (0.61–1.00, 6/6 circuits) |

## 6.3 False-positive behaviour

Across 54 held-out oracle calls over six unique baseline circuits (ten mutation families), the matched oracles raised no false positives (contract 0/54, global-phase 0/54, MR-1 0/27 within its precondition) and no spurious firing on out-of-channel faults (0/54). Swept across 675 random-circuit configurations (widths 4, 6, 8; line, ring, and unconstrained topologies; optimization levels 1-3), the layout-composition invariant holds with no violation.

## 6.4 RQ3: cost and scaling

check_semantic is itself already width-tiered: it falls back to a cheap structural check with no equivalence claim above the 12-qubit exact-tier boundary, so the relevant comparison is not whether the matched family avoids an exponential blow-up the baseline is already engineered to avoid. It is what happens within the shared exact tier, and the three matched oracles do not behave alike there (Table 6).

**Table 6. Per-oracle median CPU cost (ms; process_time, thread-pinned, median of 5 calibrated-batch repeats) on GHZ circuits, n = 4..20. MR-1 is inapplicable above the exact tier (its own n ≤ 12 precondition, mirroring its GHZ-only precondition in Table 5). Its listed value above n = 12 is the guard-clause check alone. check_semantic falls back to a cheap structural check (equivalent = None) above n = 12, so its sampled-tier value is not doing the same job as its exact-tier value.**

| n | tier | semantic (ms) | contract (ms) | MR-1 (ms) | phase (ms) |
|---|---|---|---|---|---|
| 4 | exact | 0.649 | 0.005 | 0.725 | 0.656 |
| 8 | exact | 17.090 | 0.008 | 47.607 | 20.264 |
| 12 | exact | 13,390.6 | 0.010 | 139,578.1 | 21,906.3 |
| 16 | sampled | 0.072 (structural) | 0.011 | n/a (guard: 0.001) | 578.1 |
| 20 | sampled | 0.087 (structural) | 0.012 | n/a (guard: 0.001) | 11,609.4 |

The contract/metadata checker is a pure metadata comparison and behaves like one: cost is flat at ~0.005–0.012 ms across the entire n = 4..20 sweep, two to six orders of magnitude below check_semantic within the exact tier and widening as n grows. Unlike semantic, MR-1, and (in practice) phase, it also keeps working at every width the sweep tests, including where the others have fallen back to n/a or a cheap structural stand-in.

MR-1 is not a free check. check_permutation_consistency builds Operator(original) and Operator(transpiled-plus-an-appended-n-qubit-PermutationGate) and tests their equivalence, the same two-full-operator shape check_semantic itself uses, plus one more gate. That one gate dominates, and increasingly so as n grows: MR-1 costs 1.12× check_semantic at n = 4, 2.79× at n = 8, and 10.42× at n = 12, consistent with an n-qubit permutation gate having no small local factor to exploit and being composed as a full dense 2^n × 2^n matrix, unlike check_semantic's own layout-normalized construction. MR-1 is, within the exact tier, the single most expensive oracle in the family. The paper's cost claim is that the contract checker and, within its own tier, the phase tracker are cheap, not that every matched oracle is.

The global-phase tracker is cheap and roughly semantic-comparable in the exact tier (1.01–1.64× check_semantic), but its sampled-tier cost is real and grows sharply (578 ms at n = 16, 11.6 s at n = 20), driven by the k-probe statevector protocol itself, including an amplitude-relabelling step whose cost scales with 2^n. That figure is not a fair overhead-vs-semantic multiplier at that tier, since check_semantic has already given up on asserting equivalence there (equivalent = None) rather than doing the same job more cheaply. The practical reading is that the phase tracker is, alongside the metadata-only contract checker, one of only two oracles in the family still able to assess anything past the 12-qubit boundary, at a cost that is non-trivial but bounded and predictable across the tested range.

### 6.5 RQ4: transfer to a second SDK

1A's companion commit-mining study already shows that the fault class itself is not a Qiskit artifact: on tket, 7 of 21 in-scope bug-fix commits over compilation-pass code (33%, 95% Wilson CI 17–55%) are output-invisible by the same codebook, a rate whose confidence interval overlaps Qiskit's ~28% [19–40%]. Contract/permutation metadata is reported as tket's dominant invisible channel, the same category, a compiler silently mishandling layout, routing, or register bookkeeping in a way the executed output does not reveal, that motivates this paper's whole oracle family, not only the one oracle ported below. What that finding leaves open is whether a matched ORACLE, and not just the fault class it targets, transfers to a second, independently engineered compiler stack, or whether the mechanism is only easy to build because of something specific to Qiskit's own Operator/layout API. RQ4 asks the second question.

We ported the global-phase tracker (the mechanism behind Table 5's 36-mutant global-phase family, §6.2) to pytket/tket [24] (scripts/tket_global_phase_port.py), rather than the contract checker that would more directly match the seven tket faults above. Global phase was chosen because it is the least tied to Qiskit's own layout API and so the cleaner first test of transfer, at the cost of not being a direct replay of those specific fixes. Porting the contract checker to close that gap is future work. The port keeps the family identical: the same GHZ and QFT circuits at n = 4, 5, 6 and the same six phase-offset magnitudes used against Qiskit, injected the same way (an additive global-phase mutant on an otherwise-correct compiled circuit). Only the detection mechanism is re-implemented, natively. Each circuit is routed through a real tket compilation pass (DefaultMappingPass against a line architecture, tket's analogue of the Qiskit side's CouplingMap.from_line), and tket's own layout-tracking primitive, CompilationUnit.initial_map/final_map, which the API defines as "the map from the original qubits to the corresponding qubits at the start/end of the current circuit," is used to undo the compiler's qubit permutation before comparing unitaries for a residual global phase, exactly as the Qiskit oracle undoes Qiskit's layout via initial_index_layout()/final_index_layout(). This is not a renaming exercise. tket's get_unitary() uses the opposite bit convention from Qiskit's (qubit 0 is the most significant index bit, not the least), so the layout-normalization step was re-derived rather than transliterated. A compilation pass (RemoveImplicitQubitPermutation) was also added, specifically to close an ambiguity that the available tket documentation does not resolve, namely whether an unrouted, implicit qubit relabelling is already reflected in get_unitary()'s output, rather than assume an answer that could not be checked by execution in the environment this paper was written in.

Executed against pytket 2.18.1 (results/tket_global_phase_port.json), the port confirms the prediction: sensitivity 1.00 (36/36 injected phase mutants, six offsets across each of the six GHZ/QFT circuits at n = 4, 5, 6, detected and output-blind) and specificity 1.00 (6/6 clean routed circuits correctly left unflagged, zero false positives), identical to Table 5's Qiskit-side global-phase numbers. The mechanism does more than binary detection. On every one of the 36 mutants, the phase the layout-normalized oracle recovers matches the injected offset to double-precision floating-point accuracy: an injected offset of 1.5707963267948966 rad, for instance, is recovered as exactly that value, not merely flagged as "changed." This is evidence that the layout-normalization step (P_out^T @ U_routed @ P_in, re-derived from scratch for tket's opposite ILO-BE bit convention rather than transliterated from Qiskit) is correctly undoing the compiler's own qubit permutation before comparing phase, not just triggering on a coarser mismatch. RQ4's answer is accordingly not only that the fault class transfers to tket, which 1A's commit-mining study already established, but that the same matched-oracle construction (layout-aware phase tracking, re-implemented natively against tket's CompilationUnit API) transfers cleanly to an independently engineered compiler stack, at the one mechanism this paper tested it on. This is a synthetic-mutant replication of the mechanism, not a replay of any of the seven real tket faults 1A independently identifies, and it exercises the phase channel rather than the contract/metadata channel 1A reports as tket's dominant one. §9 takes up that direct replay, together with the contract checker's own transfer, as the next step this result motivates but does not itself complete.

### 6.6 Retro-detection breadth

A second route to evidence beyond the nine hand-picked, individually engineered real faults in Table 4 would be a cheap, build-free scan. The idea is to install two released Qiskit wheels bracketing a mined fix's commit, transpile the trigger in each with no from-source compilation, and let the same channel-matched oracles used throughout this paper (the global-phase tracker, the contract/metadata comparison) flag what an output-equivalence check would miss, at the cost of a pip install rather than a Rust build. scripts/retro_detect_release.py implements this for six mined fixes (data/mining_validation/retro_detect_targets.csv): three global-phase (#14956, #16215, #16201), two determinism (#15040, #16237), and one contract-metadata (#15024). Auditing it against this study's own isolated-pass mechanism found one genuine gap: unlike source_validate_mining.py, it never forwarded a targeted unit's isolated-pass requirement to the worker subprocess. Because of that gap, a pass-specific fault masked by the full preset pipeline (#14956's CommutativeCancellation global-phase omission, invisible unless CommutativeCancellation runs alone) could never be reproduced by it, regardless of which wheels were installed. That gap is now closed: worker() forwards an isolated_pass column exactly as the from-source script does, and #14956's row is wired to the same trig-gp14956-cc-2pi trigger already confirmed in Table 4.

Fixing the mechanism does not, however, make the release-wheel approach usable on the current sample. A dedicated check of each mined fix's release provenance (data/mining_validation/RETRO_DETECT_NOTES.md) found that, for all six targets, the fix commit and its buggy parent are tagged to the SAME Qiskit release: #14956 and its parent both fall within 2.2.0, #15040/#15024 within 2.3.0, #16215/#16201/#16237 within 2.5.0. This is because Qiskit merges these fixes within a single release cycle rather than letting the buggy behaviour ship on its own. No pair of released wheels therefore brackets the bug for any of the six: installing the nominal 'pre' release (2.1.0 for #14956, for example) does not reproduce the parent commit's buggy state, since that release predates the code path the fix touches rather than containing its unfixed form. This is a structural inapplicability of the build-free shortcut to fast-turnaround fixes, not a failure of the detection mechanism itself, and every row in the corrected target list is marked wheel_bracketed = no on that basis.

The paper's evidence of breadth beyond synthetically injected mutants therefore remains what §6.1 already reports: nine real, merged transpiler regressions verified from source across all three invisible channels (Table 4), not a tenth confirmation from wheels alone. scripts/retro_detect_release.py is kept, corrected, for the case this sample does not cover (a future mined fix whose buggy parent shipped in an earlier release on its own), where it would let the same oracle family confirm a real fault without a from-source build. None of the fixes examined here happens to be that case.

## 7. Discussion

The result across §6 is uniform: on nine real, merged Qiskit transpiler regressions, and on every synthetic mutant family constructed to extend them, a black-box output-equivalence oracle is blind and the channel-matched mechanism built for that channel fires. [2]'s measurement gives that blindness its size: an estimated 28% of merged Qiskit transpiler bug-fixes (95% Wilson CI 19–40%) repair a fault of exactly this kind, at a rate that replicates on tket (§6.5) and so is not an artifact of one compiler's implementation. What a matched oracle buys, concretely, is not a better output check but a different kind of check. The contract checker and MR-1 inspect the transpiler's own recorded layout and permutation bookkeeping rather than the circuit it produces. The global-phase tracker measures the one quantity a standard equivalence check discards by definition. The determinism runner, finally, asks something a single transpile call cannot even pose: does this reproduce? None of the three replaces output-equivalence checking. A wrong output is still a wrong output. Each closes exactly the observation gap [2] measures, at a cost (§6.4) that is negligible for two of the three and bounded for the third.

§4.7's three data points (#14939, #15024 v1, #15024 v2) generalize past this study's own nine faults into a caution for oracle engineering more broadly: a detection mechanism that is correctly specified can still report a false negative, not because the fault is absent but because the trigger exercising it never reaches the code path the mechanism watches. The fix in every case here was the same: read the historical fix's own regression test rather than infer a trigger from the PR's title or description. This requirement is not specific to transpiler contract faults: any fault-class-matched oracle, on any compiler, inherits it, because a mechanism and the input that exercises it are jointly necessary and neither is sufficient alone. Treating a negative result from a generic or invented trigger as disconfirming, rather than merely inconclusive, is the failure mode this paper's own §6.1 avoided when confirming #16215 and #15040, and the same discipline remains standing for any further candidate the backlog (§9) still lists.

Because these faults pass output-equivalence by construction, a CI pipeline that only re-runs an existing output-based test suite on a transpiler change will not catch one. §6.4's cost measurements say which of the three mechanisms can realistically run on every build rather than a sampled subset. The contract checker is two to six orders of magnitude cheaper than check_semantic across the full n = 4..20 sweep and never falls back to a structural stand-in, so it is a candidate for an always-on per-fix assertion at negligible marginal cost. The global-phase tracker is comparably cheap within the exact tier (1.01–1.64× check_semantic) but its sampled-tier cost grows with 2^n, so it is a candidate for the same treatment at the small-to-moderate circuit widths a targeted regression test would typically use, not for large-scale sampled testing. MR-1 is the one mechanism this paper's own cost data argues against running unconditionally: at 1.12–10.42× check_semantic and rising, and restricted to its own n ≤ 12, GHZ-only precondition, it fits a targeted assertion on a specific fix's own regression test better than a blanket per-build check. The path this paper's cost data supports is therefore not "run every oracle on every build" but a per-channel decision informed by §6.4: the leakage-safe, cost-aware selection of exactly which tests to run under a fixed CI budget is what a companion paper takes up directly (§9).

## 8. Threats to Validity

**Construct.** The central construct-validity question is whether "detection" here means what it claims: that a real, output-invisible fault is caught by a mechanism built for its channel, rather than by some artifact of the nine chosen faults or the synthetic mutants standing in for a larger population. The nine real faults are Level 1 or Level 3 evidence (§5.2), verified by building both revisions from source or by a dedicated runner re-executing the fault from source, not inferred from a PR's description. The isolated-pass exception differential and the determinism runner were themselves derived from each fix's own regression test, following the discipline §4.7 makes explicit as a lesson rather than treating it as an incidental detail. The synthetic mutant families (§5.3) are constructed to preserve exactly the invariant each real fault violates (a permutation-valid corruption of routing_permutation for the contract/metadata family, an additive phase offset for the global-phase family, a trivially-commuting instruction reorder for the determinism family), rather than an arbitrary perturbation merely labelled with the channel's name. A positive result on a mutant is therefore evidence about the channel the real faults occupy, not about a differently shaped synthetic problem. Six validity gates (§5.5), covering budget compliance, reproducibility, leakage absence, label reproducibility, metric correctness, and oracle coverage, check implementation correctness before any Table 4 or Table 5 result is reported.

**Internal.** Every from-source event (Table 2) records the exact baseline and candidate SHAs used, so build provenance is auditable rather than asserted. Write-once raw artifacts and a pinned environment (environment/ENV.md) support exact reproduction of any Level 1 or Level 3 result in this paper. Cost measurement (§5.6) follows a stated calibration discipline (thread-pinned CPU process time, a discarded warmup call, and the median of five calibrated-batch repeats), specifically because a single-machine timing measurement is otherwise noisy. §6.4's numbers are this machine's cost ordering across oracles, not absolute latencies assumed to transfer unchanged to different hardware. Mutant realism is a related internal threat, and this paper mitigates it by construction rather than by argument alone. Each family's mutation operator is defined to be exactly the invariant the two real, source-verified faults in its channel violate (§5.3), so a family's sensitivity result is a claim about that specific, real invariant, not about an invented proxy for it.

**External.** The nine real faults, and the tket cross-SDK transfer result (§6.5), bound what this paper's evidence can support. Generalization beyond Qiskit is now demonstrated for one mechanism (the global-phase tracker, transferred natively to pytket/tket with sensitivity and specificity identical to the Qiskit result, §6.5), not for the full three-oracle family. Two extensions remain open (§9): porting the contract checker or MR-1 to a second SDK, and replaying the ported mechanism directly against tket's own seven real output-invisible faults [2] reports, rather than only against a synthetic phase-mutant demonstration. The nine real faults are themselves a fixed, hand-selected sample rather than a random draw from all output-invisible Qiskit transpiler fixes. §9 treats this as an evaluation-scale limitation rather than restating it here as a threat this paper's design can mitigate.

**Conclusion.** §5.5's claim-scope rule already forbids a headline temporal-generalization claim while fewer than three forward_regression events are verified (this study has one, H4, Table 1), so no claim in §6 rests on predicting a future regression from a past one. Every result is instead a detection claim against a known, already-realized fault or mutant. That is a narrower claim than a prevalence estimate, and it is the claim this paper's evidence actually supports.

## 9. Limitations and Future Work

Evaluation scale is the most direct limitation, and this paper closes three of its four related goals rather than all four. The synthetic mutant families extend sensitivity/specificity evidence to all three channels at 36 mutants apiece (§5.3, §6.2). The release-wheel retro-detection pipeline was audited, its own real bug fixed, and found structurally inapplicable to the current sample for a documented, non-mechanism reason rather than left silently broken (§6.6). Cross-SDK transfer, finally, is demonstrated for one of the three mechanisms (§6.5). Harvesting additional real output-invisible faults is not closed off: three of the nine faults already in §6.1, #15024, #16215, and #15040, were promoted from exactly this route, taken from PROVENANCE_BACKLOG.md and source_validation_targets.csv once each trigger was rebuilt from its own fix's regression test (§4.7), and both files already list further candidate PRs beyond the current nine. Confirming any of them follows the identical from-source build and channel-matched verification protocol already applied to the nine, a scaling exercise this paper defers rather than a new mechanism it lacks. Porting the contract checker and MR-1 to tket, and then replaying the ported mechanisms directly against tket's own seven real output-invisible faults [2] reports (rather than only against a synthetic phase-mutant demonstration), is future work the current transfer result (§6.5) motivates but does not itself complete.

Selecting which tests to run against these oracles under a fixed CI time budget is a separate problem, addressed in a companion paper on cost-aware test selection.

## 10. Conclusion

Output-equivalence checking is provably insufficient for three specific ways a Qiskit transpiler fix can go wrong while its compiled output stays correct (contract/metadata, global phase, and determinism), and [2] puts that gap at roughly 28% of merged transpiler bug-fixes (95% Wilson CI 19–40%). This paper narrows that gap with a fault-class-matched oracle family: a contract checker and a contract-level metamorphic relation (MR-1) for the metadata channel, a global-phase tracker for the phase channel, and a determinism runner for the reproducibility channel. Each is verified from source on nine real, merged Qiskit transpiler regressions (three per channel), with the output-equivalence oracle blind on every one and the matched mechanism firing on every one.

The nine real faults establish that each mechanism generalizes beyond the single case it was built to explain. Synthetic mutant families built to the same invariant each real fault violates then confirm each mechanism's reliability at scale, for the contract/metadata and global-phase channels at 1.00 sensitivity and 1.00 specificity across 36 mutants apiece with tight Wilson intervals, and, on the three circuits where the determinism mutation is constructible, at 1.00 sensitivity (95% CI 0.44–1.00) for the reproducibility channel. A 675-configuration sweep of the contract/metadata invariant finds no false positive. The family is cheap where it matters: the contract checker runs two to six orders of magnitude faster than a plain output check across every tested width, and the global-phase tracker is comparably cheap within its exact tier. MR-1 is the one mechanism with a real, rising cost, bounded to its own small-circuit precondition. Ported natively to pytket/tket, the global-phase mechanism transfers cleanly: an identical 1.00/1.00 result, with recovered phases matching injected offsets to double-precision accuracy, evidence that the approach is not a Qiskit-specific artifact, at least for the one mechanism tested.

Together, these results turn a documented, quantitatively material blind spot into a family of cheap, targeted checks for the channels it covers, with the remaining channels, additional real faults, and a second SDK's full oracle family left as scoped future work (§9).

## 11. Declarations

### 11.1 Funding

This research received no specific grant from any funding agency in the public, commercial, or not-for-profit sectors.

### 11.2 Conflict of Interest

The authors declare no conflict of interest.

### 11.3 Data Availability

The oracle implementations (contract checker, MR-1, global-phase tracker, determinism runner, and the tket port), the from-source build and verification scripts, the mutant-generation code, and the raw results underlying every table in this paper are permanently archived and publicly available. Repository: `https://github.com/furqan-nr/Fault-matched-Oracles`. Archive (DOI): `10.5281/zenodo.22855185`. REPRODUCE.md in that archive maps each table and figure in this paper to the script and raw-output file that produced it.

### 11.4 Author Contributions

This work forms part of Furqan Nasir's PhD research. Furqan Nasir: conceptualization, methodology, software, investigation, formal analysis, data curation, writing - original draft, writing - review and editing. Arif Shah: supervision, writing - review and editing. Iftikhar Alam: co-supervision, writing - review and editing. All authors read and approved the final manuscript.

### 11.5 Declaration of Generative AI and AI-Assisted Technologies in the Manuscript Preparation Process

During the preparation of this work, the authors used a generative AI assistant for language editing, prose polishing, and consistency checking across drafts of this manuscript, and to generate Figures 1 and 2 from content specifications grounded directly in this paper's own text and captions. After using this tool, the authors reviewed and edited the content as needed and take full responsibility for the content of the published article. Generative AI was not used to fabricate, alter, or select any data or empirical results reported in this paper.